# Study of SARS-CoV-2 Spike Protein by Surface Enhanced Raman Spectroscopy and Transmission Electron Microscopy

Monika Ghalawat[2], Virendra Kumar Meena[1], Sharda Prasad[1], Pankaj Poddar[2#], Atanu Basu[1]*

[1] ICMR-National Institute of Virology, 20A Ambedkar Road Pune, India

[2] CSIR-National Chemical Laboratory, Dr. Homi Bhabha Road Pune, India

#,* *Joint corresponding authors*

*Graphical abstract*

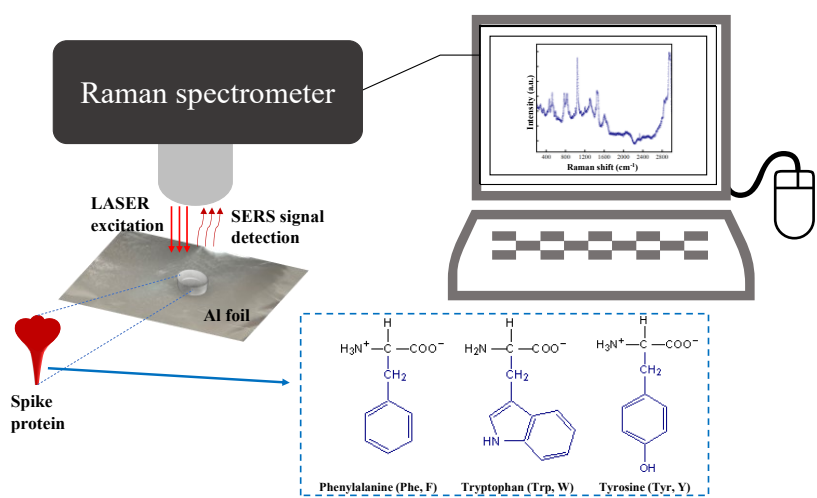

## Abstract

The spike protein (SP) of SARS-CoV-2 is the major molecular target for making diagnostic tests, vaccines, and therapeutic development. We used a combination of transmission electron microscopy (TEM) and surface enhanced Raman microscopy (SERS) to study its structure. Using SERS on an aluminum substrate, we were able to detect a characteristic spectrum of SP mostly due to vibration of three aromatic amino acids producing Raman shifts at 466 cm$^{-1}$, 524 cm$^{-1}$, 773 cm$^{-1}$, 831 cm$^{-1}$, 1048 cm$^{-1}$, 1308 cm$^{-1}$, 1457 cm$^{-1}$, and 1610 cm$^{-1}$. Transmission Electron Microscopy (TEM) of the SP showed periodic 2D-lattice orientation. The findings from this study have translational values for developing surface-enhanced Raman spectroscopy (SERS) based detectors for screening and testing SARS-CoV-2 signatures in diagnostic settings and contamination tracking.

**Corresponding Author**

E-mail: p.poddar@ncl.res.in#, basu.atanu@gov.in*

# Introduction

Alarming conditions have been generated worldwide due to the COVID-19 pandemic's rapid spread, which was brought on by the severe acute respiratory syndrome coronavirus 2 (SARS-CoV-2). When a disease manifests, it might be asymptomatic or have serious consequences, including potentially fatal respiratory tract infections. Clinical signs often arise a few days after infection.[1-3] Currently, there are two main categories that can be used to classify COVID-19 diagnostic tests. The molecular diagnostic test for the detection of SARS-CoV-2 viral RNA utilizing reverse transcriptase real-time polymerase chain reaction (RT-PCR) and nucleic acid hybridization techniques falls within the first category. Although the RT-PCR approach has emerged as the gold standard for the identification of the SARS-CoV-2 virus, it is time-consuming, involves highly skilled employees, and the risk of infection should be closely monitored throughout the procedure. Serological and immunological tests that are primarily concerned with the detection of antibodies fall under the second category. The majority of patients develop an antibody response at around 10 days following the onset of symptoms; therefore, despite the fact that these tests are quick and simple to use, they are not appropriate for screening early and asymptomatic cases. Therefore, for a prompt and correct diagnosis of viruses such as COVID-19, it is necessary to create an extremely sensitive, accurate, quick, and inexpensive diagnostic techniques. [1-3]

The SARS-CoV-2 virus as a whole and its structural proteins, including the spike protein (SP), small envelope (E) protein, and a few auxiliary proteins, might theoretically be employed as antigens for COVID-19 diagnosis. SP may be one of the most useful antigen indicators for the diagnosis of COVID-19 among them. The trimeric SP, which the virus uses to enter host cells, is the target of numerous vaccines and antibody therapy created to fight SARS-CoV-2 and lessen the severity of COVID-19 disease. It is regarded as the largest surface protein virus, a 180 kDa spike monomer, allowing it to adhere to the human host cells more quickly. [1-3] Diagnostic methods that can swiftly and sensitively identify SARS-CoV-2 and fast pivot to identify newly emerging virus variants are crucial in controlling viral spread because changes in the SP can result in the formation of virus variants with greater transmissibility. Thus, SP is essential because it acts as a mediator for the host cells through the Angiotensin-Converting Enzyme 2 (ACE-2) receptor. [1-3]

Surface-enhanced Raman spectroscopy (SERS), one of several advanced analytical tools, has demonstrated its efficiency in terms of persistent, extremely high sensitivity, and excellent selectivity for identifying biomaterials. By customizing the design of substrates, the technique has gradually progressed from the fundamental detection of chemical fingerprints of organic molecules to an alternative bio-analytical tool for identifying microbes to a greater extent with excellent specificity and reproducibility. With SERS, microbiological organisms can be quickly detected at the trace level, decreasing the spread of their different consequences on human civilization. SERS grew into a sensitive technology with a rapid detecting mechanism.[1-3]

There is a risk for false negatives in some circumstances with RT-PCR, which is based on nucleic acids, as it relies on a specific time period for evaluating the infections as positive (early stage of infection, which is defined as 6 days prior to 14 days after the onset of symptoms). The time required for obtaining results is almost 8 hours. While SERS has no time limit for detection (at any stage of infection), findings can be obtained in less than five minutes. Since it is more surface sensitive than RT-PCR, it is possible to identify the SARS-CoV-2 virus's protein moieties consciously. As a result, it can be used independently as a quick screening method to satisfy pandemic breakout demands.[1-3]

There are few reports on the detection of the SARS-CoV-2 virus using SERS. Chisanga et al.[4] created a quick and incredibly sensitive multiplexed SERS with microfluidics integration to screen several anti-spike immunoglobulin isotypes (IgG, IgA, and IgM) of SARS-CoV-2 virus using Au nanoparticles (NPs) as a substrate. In a microliter of serum samples in less than 15 minutes, Wang et al.[5] reported using a lateral flow immunosensor to detect IgG and IgM antibodies against SARS-CoV-2. Here, they targeted IgG and IgM using $SiO_2$@AuNPs@QD nanobeads functionalized with SP. In another article, Huang et al.[6] developed a fishing mode device to identify the receptor-binding domain (RBD) of the SARS-CoV-2 at low concentrations in various detecting conditions and got a sensitive SERS signal response using AuNPs as a substrate. Peng et al.[7] observed that the $Nb_2C$ and $Ta_2C$ as a SERS substrate give the capacity to precisely and sensitively detect the SARS-CoV-2 SP. In another study[1], they used a substrate made of a novel oil/water/oil (O/W/O) three-phase liquid-liquid interfaces self-assembly method, forming two layers of uniform and dense gold NP films to ensure the reproducibility and sensitivity of SERS immunoassay for the highly sensitive detection of the SARS-CoV-2 virus in untreated saliva.

After weighing their merits over other comparable approaches, the diagnostic techniques typically used in laboratories are directed toward the commercial sector. The SERS technique has several advantages over earlier approaches and has demonstrated technological development, leading to a greatly optimized and improved outcome.[1-3] The first benefit worth emphasizing is the applicability of such a combination of procedures that can make the current diagnosis error-free, more sensitive, cost-effective, and quicker. However, at the same time, using the SERS approach for commercial applications comes with a few challenges. One such example is when the analysis of a biomolecule is to be made, the surface treatment of the substrate to make it SERS active. Gold and silver are the most extensively used plasmonic NPs in the SERS substrates. Thus, this surface treatment mainly entails the fabrication of Ag/Au NPs, which requires significant skill.[1-3] It is therefore challenging to commercialize them as a substrate. This paper addresses this significant problem in the SERS approach for commercial applications.

In this work, we developed an ultrasensitive and specific Raman spectroscopy-based biosensor for detecting the SP of SARS-CoV-2. We used Raman spectroscopy to assess the effectiveness of routinely used glass slides in medical labs and hospitals with the readily accessible aluminum (Al) foil for the SP of SARS-CoV-2. Thus, the commercially available Al foil has been used to form a SERS active substrate. This Al foil approach has observed excellent uniformity and reproducibility, guaranteeing high sensitivity, stability, and repeatability of SERS.

## Materials and Methods

### SARS-CoV-2 Glycoprotein

The full-length purified spike protein (Product code REC31868-100) expressed of SARS-CoV-2 was purchased from Native Antigens Co, USA. The protein was expressed in CHO cells with a furin cleavage mutation (amino acids 682-685) and was the full-length protein from amino acids 1-1211. The protein accession database code YP-009724390.1 and the parent virus was Wuhan Hu-1 SARS-CoV-2. The observed molecular weight of the purified protein in SDS gel electrophoresis was 170 kDa.

### Raman Spectroscopy

Raman spectra were acquired at laboratory temperature (~25 °C). An HR 800 Raman spectrophotometer (Jobin Yvon, HORIBA, France) was used for taking measurements. An achromatic Czerny–Turner type monochromator (800 mm focal-length) with silver-treated mirrors equipped having a Peiltier cooled He–Ne laser light source (633 nm) operating at 20 mW was used as the illumination source. Emitted signals were captured using spectroscopy grade multi-channel CCD detectors (1024 × 256 pixels of 26 μm) under low dark current conditions (<2 x $10^{-3}$ pixels/s) with a 50x LD (Long distance) magnification. The spectra were recorded after 5-second and 10-second time intervals, respectively. Silicon reference was used for the calibration. To achieve the SERS effect, two drops of SP (2 μl each, one after the other) were dropped on Al foil substrate with the dilution of SARS-CoV-2 SP in nuclease and protease-free double distilled autoclaved water at a concentration of 0.25μg/μL, followed by air-dried at laboratory temperature for 10-15 min. Spectra were collected in the region from 200 to 3000 cm$^{-1}$ with a resolution of $\pm 1$ cm$^{-1}$.

**Transmission Electron Microscopy (TEM)**

Approximately 2μg of the SP was adsorbed on freshly carbon-coated formvar surface of a 400-mesh electron microscopy copper grid and stained with sodium phosphotungstic acid and uranyl acetate in separate experiments.[8] The negative stained protein was examined under 120 kV operating voltage in the TEM (Tecnai-12 Biotwin, FEI Co., Netherlands) under low beam current using a LaB$_6$ electron source. Images were captured using a side-mounted 2k x 2k CCD camera and analyzed using onboard Image Analysis software (Megaview III, Olympus Germany). The microscope was corrected for astigmatism before imaging the sample.

## Results and Discussion

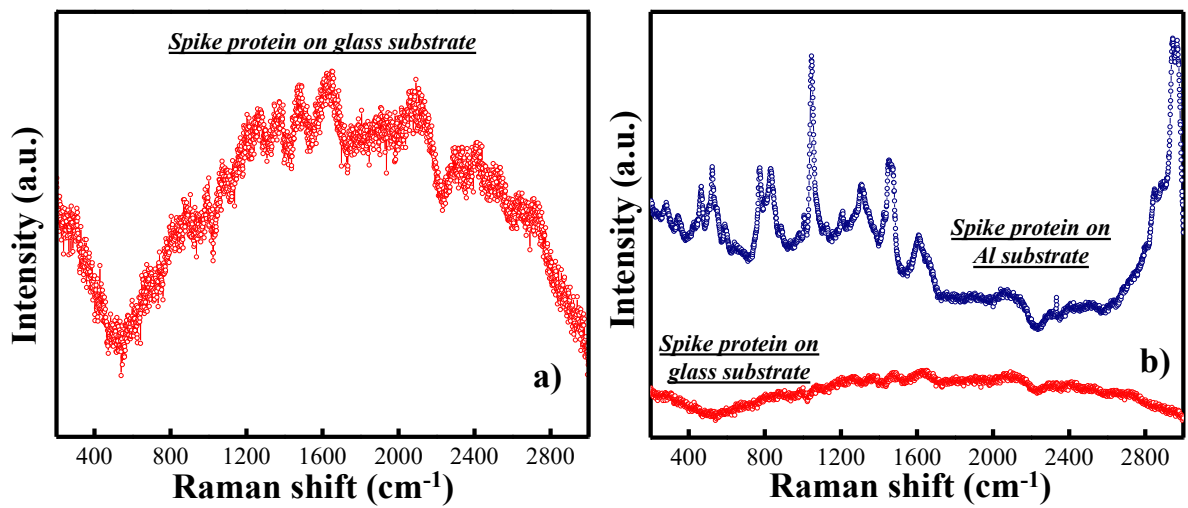

**Figure 1**. The Raman spectra of SARS-CoV-2 spike protein on the a) glass substrate and b) comparison of glass and Al substrate.

Raman spectroscopy is an ultrasensitive molecular spectroscopy technique with no interference from water, making it a distinct advantage in identifying bio-samples. For the preparation and analysis of various types of biological samples, glass substrates are frequently used in medical labs and hospitals. The SARS-CoV-2 full-length SP is first to drop cast on a glass substrate and then put under a Raman spectrometer to examine the Raman spectra. The spectra' peaks, illustrated in Figure 1a, are not at all clear. Thus, glass has been found to be unsuitable for spectroscopy because it produces background signals and distorts the biological information coming from the sample. Therefore, the best strategy would be to utilize the affordable

substrates— by applying a metallic surface on glass substrate results in significantly reducing the background noise.

The SERS approach has been utilized to improve the Raman spectra in a number of papers, typically, the Ag and Au NPs as a substrate. However, it is quite challenging to fabricate these NPs and reproduce the Raman spectra from this kind of substrate. Therefore, the high cost and skills required to fabricate Ag and Au NPs make it challenging to commercialize these substrates; therefore, we require a more affordable and simple-to-use substrate.[1-3]

The previous proof-of-concept has demonstrated the potential of Al foil as a useful substrate.[9-12] Having a strong and affordable substrate for Raman spectroscopy techniques would be very helpful. However, only very few have used this substrate for bio-spectroscopy. Here, Al foil has been used as a substrate for detecting SARS-CoV-2 SP. The SP was dropped cast on Al foil and let dry for 10 to 15 min under ambient conditions. Afterwards, the sample is placed in a Raman spectrometer and recorded the Raman spectra. Figure 1b shows the Raman spectra of SP with well-defined peaks with Al foil substrate compared to the glass substrate. Due to the Al foil's metallic composition and surface roughness, it is one of the best SERS inducers. The SERS effect induced by the Al substrate generates well-defined signal peaks and detailed spectra compared to glass (which interferes with the Raman signals). The clear spectrum produced remarkable SERS signals with variously defined peaks from 200 to 3000 $cm^{-1}$.

The fundamental components of proteins are amino acids, having a crucial physiological role in all life forms. Figure 2 shows assigned peaks of different amino acids in the Raman spectra of SP on the Al foil substrate. The highest intensity peak in the fingerprint region (800-1800 $cm^{-1}$) is 1048 $cm^{-1}$ (Figure 2). This band is assigned to phenylalanine[13-14] (deformation of the ring), tryptophan[13-15] (twisting of $NH_3^+$, stretching of benzene and pyrrole ring), and C–N and C–C protein stretching[16]. The other strongly featured Raman peaks are primarily located at 466, 524, 773, 831, 1308, 1457, 1610, 2944, and 2968 $cm^{-1}$ (Figure 2). Among these strongly visible bands, many are attributed to phenylalanine and tryptophan.[13-15] The 466 $cm^{-1}$ band appeared due to the deformation of C-C-C-C in phenylalanine and the stretching of benzene and pyrrole ring in tryptophan. Another band at 524 $cm^{-1}$ also belongs to both phenylalanine (deformation of C-C=O and ring) as well as tryptophan (twisting of CH-$CH_2$). The band at 773 $cm^{-1}$ is attributed to tryptophan (deformation of benzene and pyrrole ring, scissoring of $CO_2^-$), and the 831 $cm^{-1}$ band associated with tryptophan (twisting of $NH_3^+$, H-bending on pyrrole ring). Again, the band at 1308 $cm^{-1}$ is attributed to both phenylalanine (stretching of the ring) and tryptophan (H-bending, wagging of $CH_2$). The 1610 $cm^{-1}$ band is also present and attributed to the stretching of C=O in phenylalanine. The 1457 cm$^{-1}$ peak is featured due to the C–H stretching of glycoproteins[16]. The band observed at 831 cm$^{-1}$ in the spectra is due to the presence of tyrosine (out of plane deformation) in the protein[13,17]. This band is the marker of tyrosine. In the higher wavelength region, above 2800 cm$^{-1}$, the vibration bands are observed by virtue of the C-H stretching of aliphatic side chains[17].

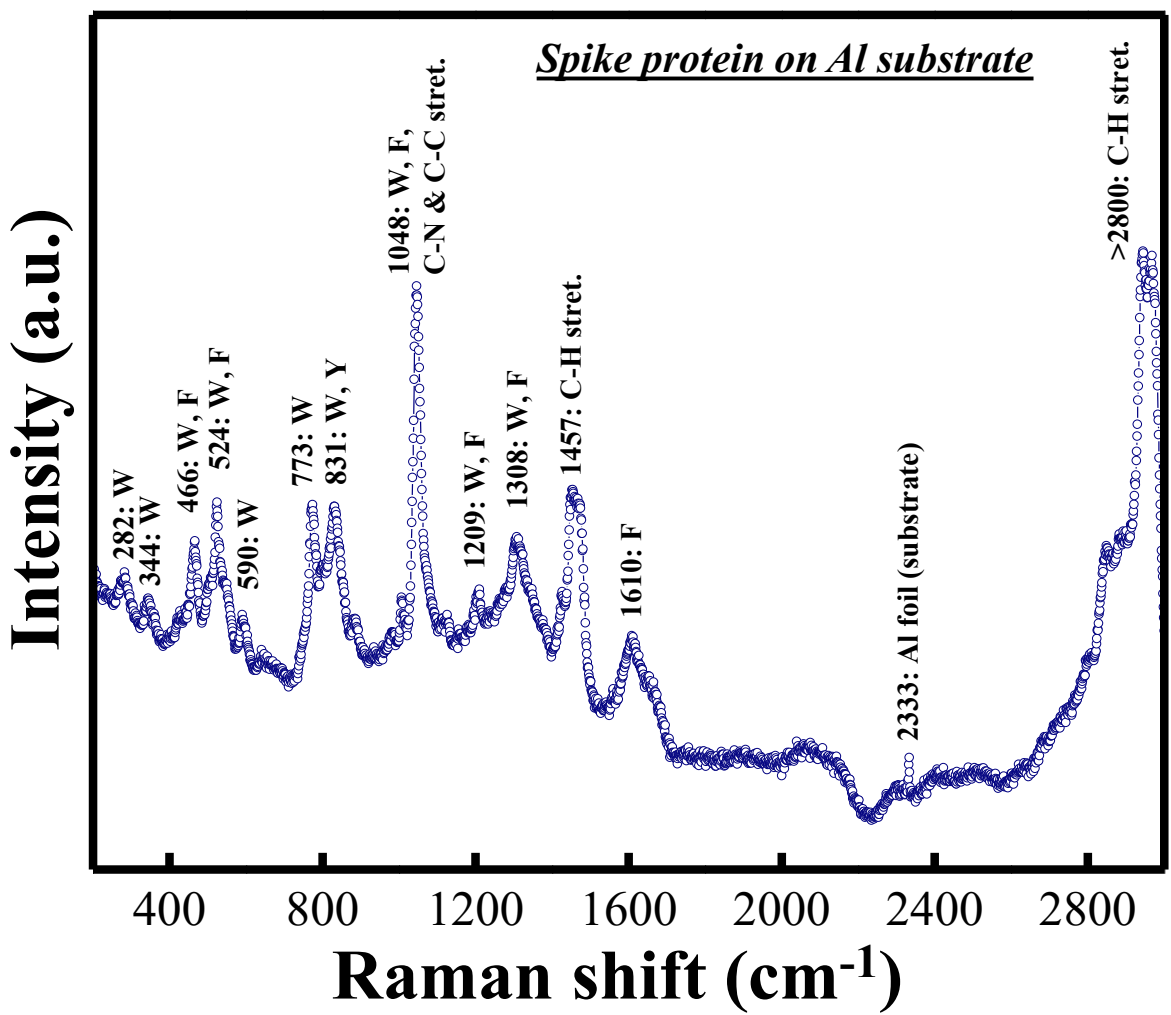

**Figure 2**. The Raman spectra of SARS-CoV-2 spike protein on Al foil substrate. The tyrosine (Tyr), tryptophan (Trp), and phenylalanine (Phe) amino acids are denoted by Y, W, and F, respectively.

The other Raman peaks detected from the SP were relatively weak and located at nearly 282, 344, 425, 590, 801, 875, 983, 1006, 1125, 1155, 1209, 1426, and 2850 cm$^{-1}$. Many bands are assigned to tryptophan[13-15]— 282 cm$^{-1}$ (rocking of CH$_2$), 344 cm$^{-1}$ (twisting of CH-CH$_2$), 425 cm$^{-1}$ (deformation of benzene and pyrrole ring), 590 cm$^{-1}$ (bending of NH in pyrrole ring as well as deformation of benzene and pyrrole ring), 801 cm$^{-1}$ (rocking of CH$_2$, stretching of C-COO⁻), 875 cm$^{-1}$ (scissoring of H benzene and pyrrole ring), 983 cm$^{-1}$ (twisting of CH$_2$, stretching of CN), 1006 cm$^{-1}$ (ring breathing in benzene and pyrrole ring), 1125 cm$^{-1}$ (H-scissoring on benzene ring, bending of CH, wagging of NH$_3^+$), 1155 cm$^{-1}$ (H-scissoring on benzene ring), 1209 cm$^{-1}$ (stretching of pyrrole ring and C-COO⁻), 1426 cm$^{-1}$ (stretching of benzene and pyrrole ring), and 2850 cm$^{-1}$ (symmetric stretching of CH$_2$, stretching CH). The following bands are assigned to phenylalanine[13-15]— 1006 cm$^{-1}$ (deformation of the ring), 1155 cm$^{-1}$ (C-H and ring deformation), and 1209 cm$^{-1}$ (out of plane bending of C-H and deformation of the ring). The band between 880 to 920 cm$^{-1}$ is also attributed to the CH$_2$ deformation[17]. The summary of the spectral signals and molecular relevance is tabulated in Table 1.

**Table 1.** The assigned peaks of SERS spectra of the SARS-CoV-2 spike protein.

| Raman peak (cm$^{-1}$) | Strength | Relevance | Reference |
|---|---|---|---|
| 466 | strong | Phenylalanine deformation and tryptophan benzene-pyrrole stretching | [13-15] |
| 524 | strong | Deformation of C-C=O and ring in phenylalanine Tryptophan (CH-CH$_2$ twisting) | [13-15] |
| 773 | strong | Tryptophan (deformation of benzene, pyrrole ring, scissoring of CO$_2^-$) | [13-15] |
| 831 | strong | Tryptophan (twisting of NH$_3^+$, H-bending on pyrrole ring), Deformation of tyrosine | [13-15] |
| 1308 | strong | Phenylalanine (ring stretching), tryptophan (H-bending, CH$_2$ wagging) | [13-15] |
| 1457 | strong | Glycoprotein CH stretch | [16] |
| 1610 | strong | Phenylalanine C=O stretch | [13-15] |
| 1048 | strongest | Phenylalanine (ring deformation), tryptophan (NH$_3^+$ twisting, benzene and pyrrole ring stretch), and Glycoprotein C–N and C–C stretch | [13-16] |
| | | | |
| 282 | weak | Tryptophan (CH$_2$ rocking) | [13-15] |
| 344 | weak | Tryptophan (CH-CH$_2$ twisting) | [13-15] |
| 425 | weak | Tryptophan (Benzene and pyrrole ring deformation) | [13-15] |
| 590 | weak | Tryptophan (Bending of NH in pyrrole and deformation of benzene and pyrrole) | [13-15] |
| 801 | weak | Tryptophan (CH$_2$ rocking and C-COO⁻ stretching) | [13-15] |
| 875 | weak | Tryptophan (Scissoring of H benzene and pyrrole ring) | [13-15] |
| 983 | weak | Tryptophan (Twisting of CH$_2$ and CN stretching) | [13-15] |
| 1006 | weak | Tryptophan (Benzene and pyrrole ring breathing) Phenylalanine (ring deformation) | [13-15] |

| 1125 | weak | Tryptophan (H-scissoring in benzene ring, CH bending and wagging of $NH_3^+$) | [13-15] |
| 1155 | weak | Tryptophan (H-scissoring of benzene ring) Phenylalanine (CH and ring deformation) | [13-15] |
| 1209 | weak | Tryptophan (Pyrrole ring and C-COO$^-$ stretching) Phenylalanine (CH bending and ring deformation) | [13-15] |
| 1426 | weak | Tryptophan (Stretching of benzene and pyrrole ring) | [13-15] |
| 2850 | weak | Tryptophan (Symmetric stretching of -$CH_2$, stretching of CH) | [13-15] |

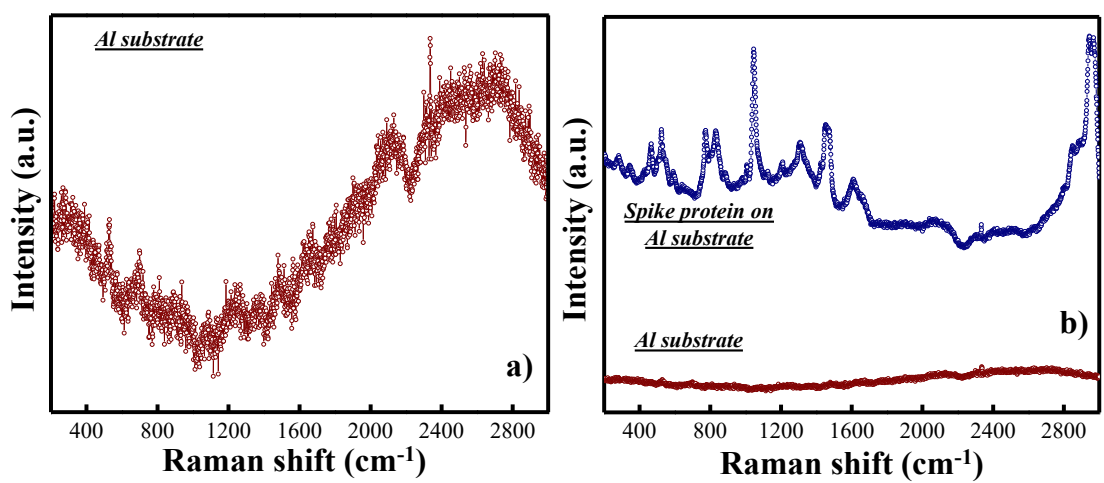

**Figure 3.** a) The Raman spectra of Al foil substrate. b) Comparison of the Raman spectra of Al foil substrate and SARS-CoV-2 spike protein on the Al foil substrate.

The above discussion clearly shows that the peaks observed in the spectra are mainly assigned to 3 types of amino acids— Tyrosine (Tyr; Y), Tryptophan (Trp; W), and Phenylalanine (Phe; F). The only three amino acids with benzyl-based aromatic groups are phenylalanine, tyrosine, and tryptophan, which are known as aromatic amino acids (AAA). The polarizability of each amino acid has to be considered, as amino acids contribute to comprehending the proteins' significance. The three AAA have the highest degree of polarizability, as reported by Millefiori et al.[18] These amino acids overshadow the contribution of the other amino acids. The position of 3 AAA in the spike protein is highlighted in Figure S1 with different corresponding colors. The 54 tyrosine (Tyr; Y), 12 tryptophan (Trp; W), and 77 phenylalanine (Phe; F) contribute to the final Raman spectra of SP. It is important to note that there is a negligible contribution in the final spectra from the Al foil. The Raman spectra of Al foil shown in Figure 3a reveal no well-defined peak, and the comparison in Figure 3b reveals that only one minor peak (~ 2333 cm$^{-1}$) is observed above the fingerprint region due to Al foil. Additionally, the contribution from the water used to dilute the SP is also observed. Figures S2a and b clearly reveal that the contribution is negligible and that the obtained spectra are purely due to the SP.

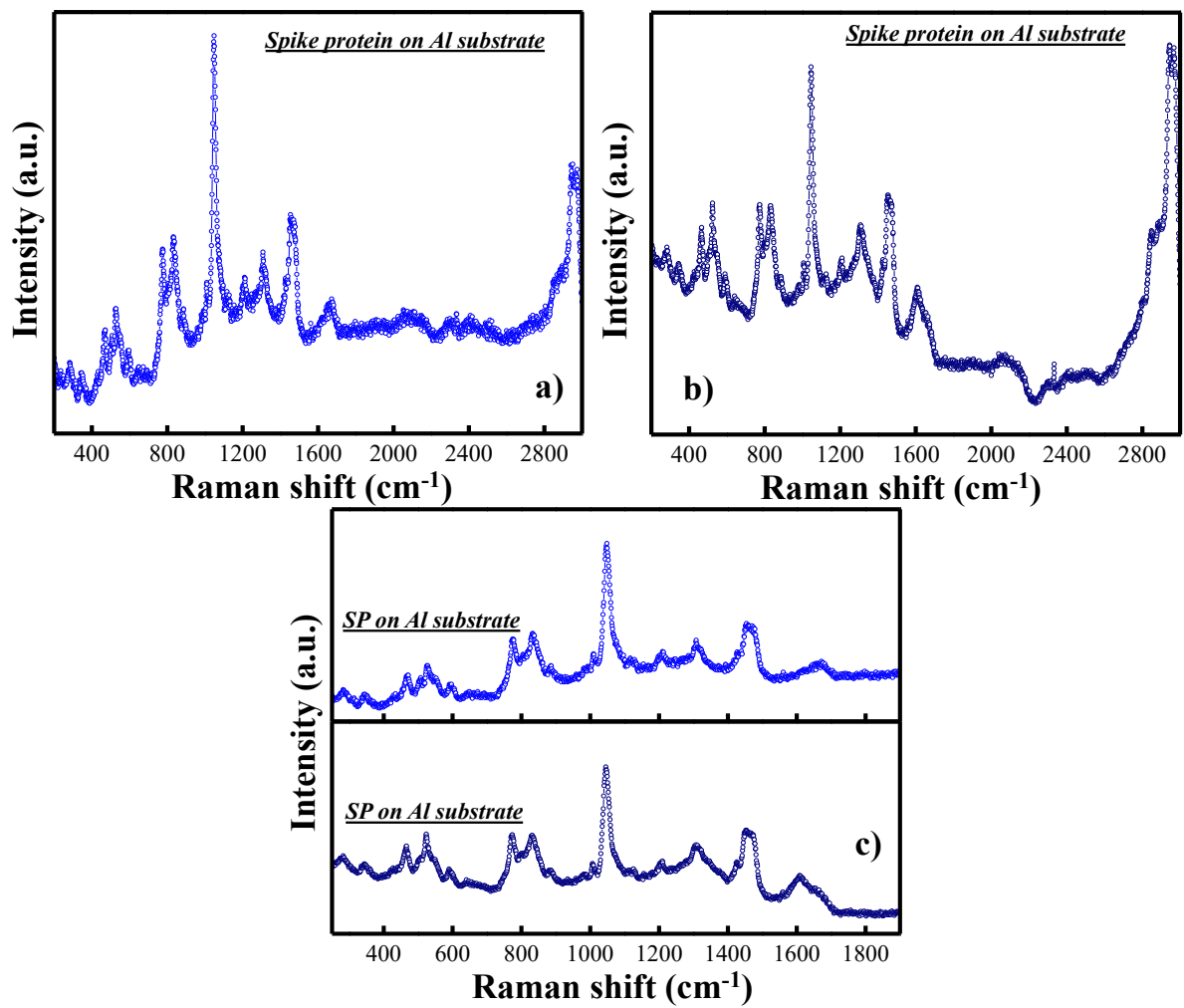

**Figure 4**. a) and b) The Raman spectra of SARS-CoV-2 spike protein (SP) on Al foil substrate clearly show the reproducibility of Raman spectra using Al foil substrate. c) Comparison of Raman spectra of SP in the fingerprint region.

Furthermore, one more important aspect to discuss is the reproducibility of the Raman spectra. SERS has been criticized for having poor reproducibility in actual samples. Figure 4 shows the reproducibility of SERS spectra of SP on Al foil substrate. The spectra in Figures 4a and 4b reveal the reproducibility with considerable variation in background noise. Figure 4c illustrates the reproducibility of Raman spectra of SP in the fingerprint region. Thus, it is appropriate to conclude that the Raman signals using the SERS method with Al foil as a substrate showed a well-defined spectrum from the SARS CoV-2 SP having high reproducibility. The observed bands are attributed to phenylalanine, tryptophan, and tyrosine.

All three AAA contain the benzene structure, and the most intense bands are present due to ring vibrations.[19] Thus, the limitations of the Raman technique in biological samples are overcome by using Al foil substrate in SERS.

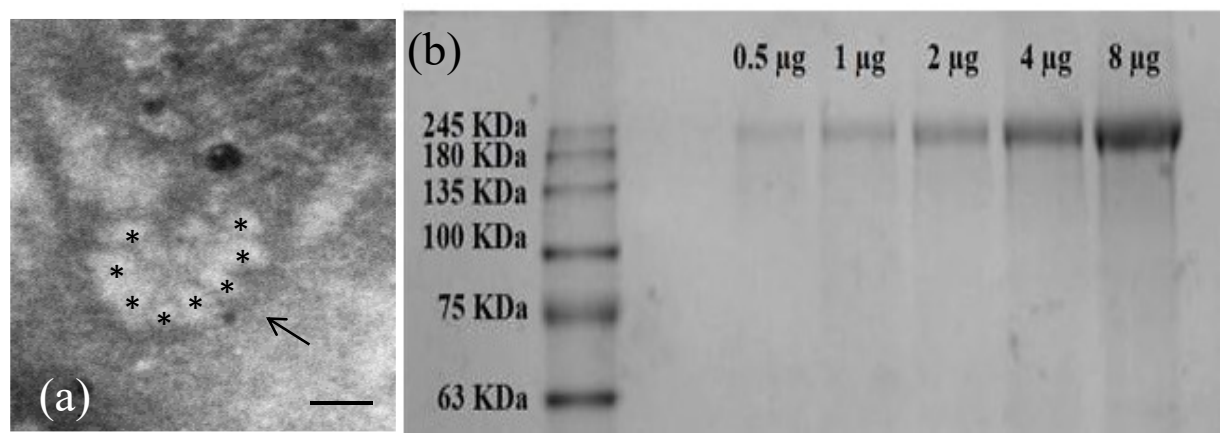

**Figure 5**. a) Representative TEM images of the spike protein described in the text. b) SDS-PAGE profile of the purified spike protein at different amounts loaded.

Negatively stained TEM imaging of the protein showed the presence of non-specific 2D lattice-like aggregates shown in Figure 5. These structures were mostly observed as stacked whorl-like arrangements in some fields or random clusters, as in Figure 5. Occasional tetrameric aggregates could also be imaged in some of the fields scanned. The average size of the subunits ranged from 8±2 nm across and the aggregates approximately 30±3 nm in length. The limitation of negative staining TEM is a relatively low resolution and presence of imaging artifacts. However, while high-resolution structure analysis was not within the scope of the present study, we used TEM for qualitative imaging of the purified protein in parallel with Raman spectral analysis to maintain consistency of the preparation. The TEM observation on the protein was consistent with the overall morphology of the SARS CoV-2 SP as reported from cryo-EM analysis, thus providing credibility to the quality of the SERS signature.[20]

**Conclusion**

In summary, the full-length SARS-COV-2 SP was observed and analyzed using SERS at 200-3000 cm$^{-1}$, and the characteristic spectrum was mapped. The Raman signals using the SERS method with Al foil as a substrate showed well-defined and consistently reproducible SP spectra. The observed bands were attributed to three AAA— phenylalanine, tryptophan, and tyrosine. All three AAA contain the benzene structure, and the most intense bands are present due to ring vibrations. Thus, it has been shown that Al foil is a highly useful material for creating a robust and cost-effective substrate for Raman spectroscopy techniques. The finding can be very informative for an advanced understanding of the molecular basis of SP interaction with ligands and has the potential for developing novel therapeutics. Another significant translational aspect of the study could be fundamental in developing high throughput non-invasive Raman-based medical devices for screening the virus, such as COVID-19, and detecting surface contamination.

**Conflict of interest**

None

**Acknowledgement**

The authors would like to acknowledge Indian Council of Medical Research (ICMR) and Council of Scientific and Industrial Research (CSIR) for supporting core instrumental facilities. One of the authors M.G has been supported by a Senior Research Fellowship Grant from the University Grants Commission. The authors would like to thank all staff of the core facilities for providing technical support.

# Supporting Information

# Study of SARS-CoV-2 Spike Protein by Surface Enhanced Raman Spectroscopy and Transmission Electron Microscopy

Monika Ghalawat[2], Virendra Kumar Meena[1], Sharda Prasad[1], Pankaj Poddar[2#], Atanu Basu[1]*

[1] *ICMR-National Institute of Virology, 20A Ambedkar Road Pune, India*
[2] *CSIR-National Chemical Laboratory, Dr. Homi Bhabha Road Pune, India*
*#,* Joint corresponding authors*

**Corresponding Author**

E-mail: p.poddar@ncl.res.in[#], basu.atanu@gov.in*

```
   1 MFVFLVLLPL VSSQCVNLTT RTQLPPAYTN SFTRGVYYPD KVFRSSVLHS TQDLFLPFFS
  61 NVTWFHAIHV SGTNGTKRFD NPVLPFNDGV YFASTEKSNI IRGWIFGTTL DSKTQSLLIV
 121 NNATNVVIKV CEFQFCNDPF LGVYYHKNNK SWMESEFRVY SSANNCTFEY VSQPFLMDLE
 181 GKQGNFKNLR EFVFKNIDGY FKIYSKHTPI NLVRDLPQGF SALEPLVDLP IGINITRFQT
 241 LLALHRSYLT PGDSSSGWTA GAAAYYVGYL QPRTFLLKYN ENGTITDAVD CALDPLSETK
 301 CTLKSFTVEK GIYQTSNFRV QPTESIVRFP NITNLCPFGE VFNATRFASV YAWNRKRISN
 361 CVADYSVLYN SASFSTFKCY GVSPTKLNDL CFTNVYADSF VIRGDEVRQI APGQTGKIAD
 421 YNYKLPDDFT GCVIAWNSNN LDSKVGGNYN YLYRLFRKSN LKPFERDIST EIYQAGSTPC
 481 NGVEGFNCYF PLQSYGFQPT NGVGYQPYRV VVLSFELLHA PATVCGPKKS TNLVKNKCVN
 541 FNFNGLTGTG VLTESNKKFL PFQQFGRDIA DTTDAVRDPQ TLEILDITPC SFGGVSVITP
 601 GTNTSNQVAV LYQDVNCTEV PVAIHADQLT PTWRVYSTGS NVFQTRAGCL IGAEHVNNSY
 661 ECDIPIGAGI CASYQTQTNS PRRARSVASQ SIIAYTMSLG AENSVAYSNN SIAIPTNFTI
 721 SVTTEILPVS MTKTSVDCTM YICGDSTECS NLLLQYGSFC TQLNRALTGI AVEQDKNTQE
 781 VFAQVKQIYK TPPIKDFGGF NFSQILPDPS KPSKRSFIED LLFNKVTLAD AGFIKQYGDC
 841 LGDIAARDLI CAQKFNGLTV LPPLLTDEMI AQYTSALLAG TITSGWTFGA GAALQIPFAM
 901 QMAYRFNGIG VTQNVLYENQ KLIANQFNSA IGKIQDSLSS TASALGKLQD VVNQNAQALN
 961 TLVKQLSSNF GAISSVLNDI LSRLDKVEAE VQIDRLITGR LQSLQTYVTQ QLIRAAEIRA
1021 SANLAATKMS ECVLGQSKRV DFCGKGYHLM SFPQSAPHGV VFLHVTYVPA QEKNFTTAPA
1081 ICHDGKAHFP REGVFVSNGT HWFVTQRNFY EPQIITTDNT FVSGNCDVVI GIVNNTVYDP
1141 LQPELDSFKE ELDKYFKNHT SPDVDLGDIS GINASVVNIQ KEIDRLNEVA KNLNESLIDL
1201 QELGKYEQYI KWPWYIWLGF IAGLIAIVMV TIMLCCMTSC CSCLKGCCSC GSCCKFDEDD
1261 SEPVLKGVKL HYT
```

**Figure S1.** The sequence of SARS-CoV-2 full-length spike protein from protein accession database code YP-009724390.1. The tyrosine (Tyr; Y), tryptophan (Trp; W), and phenylalanine (Phe; F) amino acids at distinct positions are highlighted by pink, blue and yellow, respectively. [**Ref**. https://www.ncbi.nlm.nih.gov/protein/YP_009724390.1]

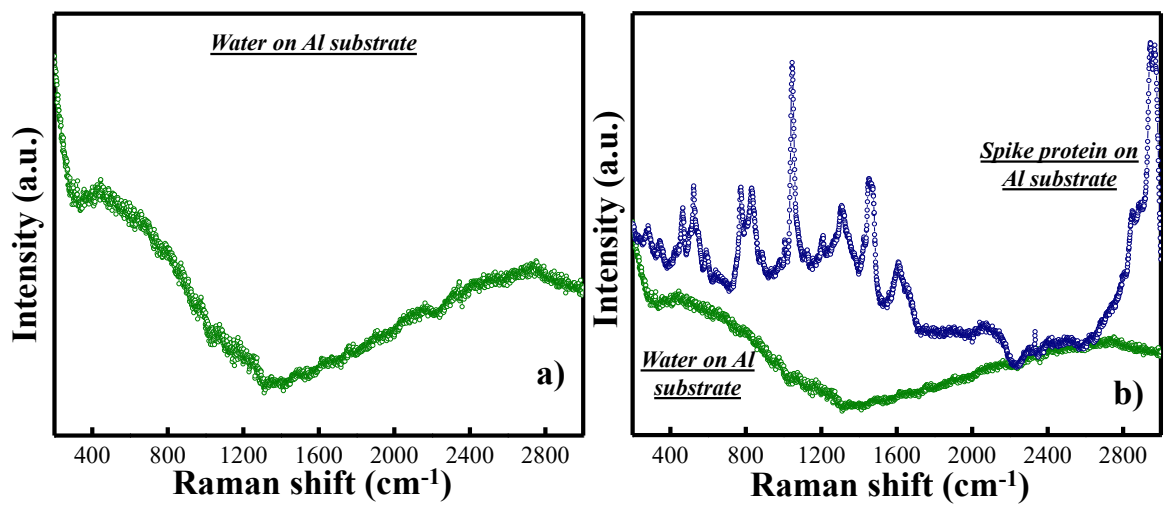

**Figure S2.** a) The Raman spectra of water on Al foil substrate. b) Comparison of the Raman spectra of water and SARS-CoV-2 spike protein on the Al foil substrate.